\documentclass[12pt]{article}
\usepackage{amssymb,amsmath}
\usepackage[noblocks]{authblk}
\usepackage[top=0.75in, bottom=0.75in, left=0.75in, right=0.75in, dvips]{geometry}
\usepackage{caption}
\begin{document}
\textwidth 10.0in 
\textheight 9.0in 
\topmargin -0.60in
\title{The Canonical Structure of the Superstring Action}
\author[1]{F.A. Chishtie}
\affil[1] {Department of Space Science, Institute of Space Technology, Islamabad 44000, Pakistan} 
\author[2,3]{D.G.C. McKeon}
\affil[2] {Department of Applied Mathematics, The
University of Western Ontario, London, ON N6A 5B7, Canada} 
\affil[3] {Department of Mathematics and
Computer Science, Algoma University, Sault St.Marie, ON P6A
2G4, Canada}
\date{}
\maketitle

\maketitle
\noindent
\noindent
email: dgmckeo2@uwo.ca, fachisht@uwo.ca

\begin{abstract}
We consider the canonical structure of the Green-Schwarz superstring in $9 + 1$ dimensions using the Dirac constraint formalism; it is shown that its structure is similar to that of the superparticle in $2 + 1$ and $3 + 1$ dimensions.  A key feature of this structure is that the primary Fermionic constraints can be divided into two groups using field-independent  projection operators; if one of these groups is eliminated through use of a Dirac Bracket (DB) then the second group of primary Fermionic constraints becomes first class.  (This is what also happens with the superparticle action.)  These primary Fermionic first class constraints can be used to find the generator of a local Fermionic gauge symmetry of the action.  We also consider the superstring action in other dimensions of space-time to see if the Fermionic gauge symmetry can be made simpler than it is in $2 + 1$, $3 + 1$ and $9 + 1$ dimensions.  With a $3 + 3$ dimensional target space, we find that such a simplification occurs. We finally show how in five dimensions there is no first class Fermionic constraint.
\end{abstract}

\section{Introduction}

The fields in superstring theory [1-3] serve as a map from a $1 + 1$ dimensional ``world sheet'' to a $d = s + t$ dimensional ``target space''  whose metric is $\eta^{\mu\nu} = \rm{diag} ( + \ldots +, - \ldots -)(s+, t-)$.  The original ``spinning string'' action has manifest local supersymmetry on the world sheet [4, 5] while the ``superstring'' action has manifest local Fermionic symmetry in the target space--the so-called $\kappa$ symmetry [6].

Local gauge symmetries are generally taken to be a consequence of the presence of first class constraints in the action, using Dirac's approach to canonical analysis [7, 8].  The generator of a local gauge symmetry can be constructed from these first class constraints using either the approach of Castellani (C) [9] or of Henneaux, Teitelboim and Zanelli (HTZ) [10, 11].  (This has been done for both the spinning particle [12] and the superparticle [13].)  There has been a discussion of the canonical structure of the superstring action with a $9 + 1$ dimensional target space [14, 15], but in this analysis, the way in which the second class primary Fermionic constraints have been projected out [15] results in a Dirac Bracket (DB) which has the property that when one takes the DB of one of these secondary constraints with any dynamical variable, it vanishes only if the constraints themselves vanishes (ie, it is only a ``weak'' equation). Normally one requires that the DB of any second class constraint with any dynamical variable vanish [7, 8] identically (ie, it is a ``strong'' equation), and thus we are motivated to reexamine the constraint structure of the superstring action.  The analysis of the constraint structure of the superparticle action suggests that it is possible to divide the primary Fermionic constraints into two parts; upon defining a DB to eliminate one of these two parts, the other part becomes first class when this DB is employed [13, 16].  We find that this is in fact what happens with the superstring action when the target space is $2 + 1$, $3 + 1$ and $9 + 1$ dimensional.  (We examine a variety of dimensions for the target space as spinors have properties that are dependent on dimension.)

The local gauge symmetries that follow from the first class constraints present in the superparticle action can be generated by the first class constraints present; these are related to the local $\kappa$-supersymmetry transformation that leaves this action invariant [13].  Similarly, the first class constraints that are present in the superstring action once the second class constraints are eliminated using a DB can be used to find a local gauge supersymmetry for the superstring theory. We illustrate this with the superstring in $2 + 1$ dimensions.  This supersymmetry is quite complicated and so we investigate if there are any dimensions for the target space in which the supersymmetry is simplified.  It is found that in $3 + 3$ dimensions, the target space supersymmetry becomes much simpler. Also, in five dimensions, no first class Fermionic constraint arises. 

We do not consider the possible contribution of an action for the metric on the world sheet to the canonical structure of the superstring action (as has been done for the Bosonic string [17]).

The conventions we use are explained in the appendices.

\section{The Superstring}

The Green-Schwarz superstring in a $9 + 1$ dimensional target space has the action [6]
\begin{align}
S = \int d\tau d\sigma  \bigg[ \frac{1}{2} & h^{ab}Y^A_a Y_{Ab} - \epsilon^{ab} x_{,a}^A \left( \overline{\theta}_{,b}^1 \tilde{B}_A \theta^1 - 
\overline{\theta}_{,b}^2 \tilde{B}_A \theta^2 \right)\nonumber \\
&+ \epsilon^{ab} \overline{\theta}_{,a}^1 \tilde{B}^A \theta^1  
\overline{\theta}_{,b}^2 \tilde{B}_A \theta^2  \bigg].
\end{align}

The world sheet metric is $g_{ab}$ with $h^{ab} = \sqrt{-g}\,g^{ab}$ so that
\begin{equation} 
det \,h^{ab} = (\sqrt{-g})^2\left(g^{00} g^{11} - (g^{01})^2\right) = -1.
\end{equation}
The world sheet coordinates are $(\sigma , \tau) (a = 1,2)$.  There are 10 Bosonic functions $x^A(\sigma ,\tau) (A = 1 \ldots 10)$ as well as two Fermionic functions $\theta^I(\sigma ,\tau) (I = 1, 2)$  defined on the world sheet, where $x^A$ is a vector and $\theta^I$ is a 
Majorana-Weyl spinor (ie, $\theta^I = C\overline{\theta}^{IT}$ and $\theta^I = B^{11} \theta^I$) on the 10 dimensional ``target space''. We have also defined 
\begin{equation}
Y_a^A = x_{,a}^A + \overline{\theta}_{,a}^1 \tilde{B}^A \theta^1 + \overline{\theta}_{,a}^2 \tilde{B}^A \theta^2 . 
\end{equation}
The Lagrangian of eq. (1) is Hermitian if we use the conventions of appendix A.

The canonical momenta conjugate to $h^{ab}$, $x^A$, $\overline{\theta}^1$ and $\overline{\theta}^2$ are given by
\begin{subequations}
\begin{align}
I\!\!P_{ab} &= 0\\
p_A &= h^{00}Y_{A0} + h^{01} Y_{A1} - \left(\overline{\theta}_{,1}^1 \tilde{B}_A \theta^1 - \overline{\theta}_{,1}^2 \tilde{B}_A \theta^2 \right)\\
\pi^1 &= \left(p^A + x_{,1}^A + \overline{\theta}_{,1}^1 \tilde{B}^A \theta^1\right) \tilde{B}_A \theta^1 \\
\pi^2 &= \left(p^A - x_{,1}^A - \overline{\theta}_{,1}^2 \tilde{B}^A \theta^2\right) \tilde{B}_A \theta^2 
\end{align}
\end{subequations}
respectively.

Eqs. (4c,d) are a pair of primary Fermionic constraints 
\begin {subequations}
\begin{align}
\chi^1 &= \pi^1 - \left( p^A + x_{,1}^A + \overline{\theta}_{,1}^1 \tilde{B}^A \theta^1\right) \tilde{B}_A \theta^1\\
\chi^2 &= \pi^2 - \left( p^A - x_{,1}^A - \overline{\theta}_{,1}^2 \tilde{B}^A \theta^2\right) \tilde{B}_A \theta^2 .
\end{align}
\end {subequations}

We now can compute the canonical Hamiltonian 
\begin{align}
\mathcal{H}_C & = h_{,0}^{ab} I\!\!P_{ab} + x_{,0}^A\, p_A + \overline{\theta} _{,0}^1 \pi^1 + \overline{\theta} _{,0}^2 \pi^2 - \mathcal{L} \nonumber \\
&\;\;\;=\frac{H}{2} \left(\Pi_A^2 + y_A^2 \right) + H^1 \Pi_A y^A
\end{align}
where we have used eq. (2) and made the definitions
\begin{subequations}
\begin{align}
H & \equiv \frac{1}{h^{00}}\\
H^1 & \equiv \frac{-h^{01}}{h^{00}}\\
\Pi_A & \equiv p_A  + \overline{\theta}_{,1}^1 \tilde{B}_A \theta^1 - 
\overline{\theta}_{,1}^2 \tilde{B}_A \theta^2\\
y^A  &\equiv Y_1^A \;\;.
\end{align}
\end{subequations}
By eq. (4a), the momenta $I\!\!P$, $I\!\!P_1$ conjugate to $H$, $H^1$ both vanish and so by eq. (6) we are led to the secondary constraints
\begin{subequations}
\begin{align}
\Sigma_s &= \frac{1}{2}\left( \Pi_A^2 + y_A^2 \right) =\frac{1}{2} \left(p_A^2 + x_{A,1}^2 \right) -\overline{\theta}_{,1}^1 \left( \chi^1 - \pi^1\right) + \overline{\theta}_{,1}^2 \left(\chi^2 - \pi^2 \right)\\
\Sigma_p &= \Pi_A  y^A = p_A x_{,1}^A  - \overline{\theta}_{,1}^1 \left( \chi^1 - \pi^1\right) - \overline{\theta}_{,1}^2 \left(\chi^2 - \pi^2 \right)
\end{align}
\end{subequations}

By eq. (A.39) we find that 
\[ \theta^I = C\overline{\theta}^{IT}, \quad \theta^{IT} = -\overline{\theta}^IC, \quad \overline{\theta}^{IT} =  C\,\theta^I, \quad \overline{\theta}^I = - \theta^{IT}C\eqno(9a-d) \]
and so the canonical Poisson Brackets (PB) of eqs. (B.5-7)
\[ \left\lbrace x^A, p_B \right\rbrace = \delta_B^A \eqno(10a)\]
\[\hspace{.5cm}\left\lbrace \pi^I, \overline{\theta}^J \right\rbrace = -\delta^{IJ} \eqno(10b)\]
imply that
\[ \left\lbrace \chi^1, \overline{\chi}^2 \right\rbrace = 0\]
\[ \left\lbrace \chi^1, \overline{\chi}^1 \right\rbrace = -2\left(p^A + x_{,1}^A + \overline{\theta}_{,1}^1 \tilde{B}^A\theta^1\right) \tilde{B}_A + 2 \left(\tilde{B}^A\theta_{,1}^1  \overline{\theta}^1 \tilde{B}_A - 
\tilde{B}^A \theta^1  \overline{\theta}_{,1}^1 \tilde{B}_A \right) \eqno(11a) \]
\[ \hspace{.35cm}\left\lbrace \chi^2, \overline{\chi}^2 \right\rbrace = -2\left(p^A - x_{,1}^A - \overline{\theta}_{,1}^2 \tilde{B}^A\theta^2\right) \tilde{B}_A  - 2 \left(\tilde{B}^A\theta_{,1}^2  \overline{\theta}^2 \tilde{B}_A - 
\tilde{B}^A \theta^2  \overline{\theta}_{,1}^2 \tilde{B}_A \right). \eqno(11b) \]

Furthermore, if we use test functions $f(\sigma), g(\sigma)$ as in ref. [18] we find that 
\[ \hspace{-2.5cm}\int d\sigma \,d\sigma^\prime f(\sigma) \left\lbrace\Sigma_p(\sigma), \Sigma_p(\sigma^\prime)\right\rbrace g(\sigma^\prime) \nonumber \]
\[ = \int d\sigma \,d\sigma^\prime f(\sigma) \left\lbrace\Sigma_s(\sigma), \Sigma_s(\sigma^\prime)\right\rbrace g(\sigma^\prime) \nonumber \]
\[ = \int d\sigma \Sigma_p \left( fg_{,1} - f_{,1} g\right) \eqno(12a) \]
\[ \int d\sigma \,d\sigma^\prime f(\sigma) \left\lbrace \Sigma_s(\sigma), \Sigma_p(\sigma^\prime)\right\rbrace g(\sigma^\prime)= \int d\sigma \left[ \Sigma_s
\left( fg_{,1} - f_{,1}g\right) + \frac{1}{2} \left( fg(\Pi^2 - y^2)\right)_{,1} \right] \eqno(12b) \]
\[ \int d\sigma \,d\sigma^\prime f(\sigma) \left\lbrace \Sigma_p(\sigma), \Sigma_s(\sigma^\prime)\right\rbrace g(\sigma^\prime)= \int d\sigma \left[ \Sigma_s
\left( fg_{,1} - f_{,1}g\right) - \frac{1}{2} \left( fg(\Pi^2 - y^2)\right)_{,1} \right] \eqno(12c) \]
\[\int d\sigma\, d\sigma^\prime \; f(\sigma) \left\lbrace\chi^I(\sigma), \Sigma_{(s,p)} (\sigma^\prime)\right\rbrace g(\sigma^\prime) = 0.\eqno(13)\]

We now must classify the constraints $\chi^I$ as being first or second class. This problem has been addressed using field dependent operators that become projection operators only on the constraint surface, and which result in a DB which is not strongly equal to zero when the DB of a second class constraint is taken with any other dynamical variable [15].  It may be possible to circumvent this whole problem by re-expressing the superstring action in terms of twistors [19] or by using harmonic superspace [24].  We will use an approach employed when examining the canonical structure of the superparticle action [13].  Although this method lacks manifest covariance, it does lead to a separation of the Fermionic constraints into two groups; if either of the groups is treated as being second class and the appropriate DB is defined to eliminate it, then the remaining group becomes first class with this DB.

We shall simplify our discussion by first considering the $N = 1$ limit of the superstring action of eq. (1); that is we will set $\theta^1 = \theta$ and $\theta^2 = 0$.  Next, we will consider the superstring action in $2 + 1$ and $3 + 1$ dimensions before examining it in $9 + 1$ dimensions.  Remarkably, despite the differing properties of spinors in $2 + 1$, $3 + 1$ and $ 9 + 1$ dimensions, eqs. (1-8, 10-13) are all valid in each of these dimensions in the $N = 1$ limit. 

In $2 + 1$ dimensions when $\theta = C\overline{\theta}^T$, eq. (11a) when combined with the Fierz identity of eq. (A.56) results in 
\[\left\lbrace \chi , \overline{\chi} \right\rbrace = -2 \left( p^i + x_{,1}^i + 2\overline{\theta}_{,1} \tilde{\tau}^i \theta \right) \tilde{\tau}_i\nonumber \]
\[ = -2 \left( \Pi^i + y^i \right)\tilde{\tau}_i \eqno(14) \]
where now
\[\Pi^i \equiv p^i + \overline{\theta}_{,1} \tilde{\tau}^i \theta \eqno(15a) \]
\[\hspace{.4cm} y^i \equiv x_{,1}^i + \overline{\theta}_{,1} \tilde{\tau}^i \theta\,. \eqno(15b) \]
Since
\[ \left[ -2 (\Pi^i + y^i)\tilde{\tau}_i\right]^{-1} = -\frac{1}{4} 
(\Pi^i + y^i)\tilde{\tau}_i / \left[ \Sigma_s + \Sigma_p \right] \eqno(16) \]
it follows that the DB defined using $\chi$ is ill defined on the constraint surface $\Sigma_s = \Sigma_p = 0$.  To circumvent this difficulty, we are forced to follow the approach used when considering the superparticle [13] and look at the two components of $\chi$ separately.  This could be done using the projection operators $P_{\pm} = \frac{1}{2}( 1 \pm \sigma^3)$, but instead we use the components of $\theta$ explicitly. If 
\[ \theta = \left( \begin{array}{c} u \\ d \end{array} \right) \eqno(17) \]
then
\[ \overline{\theta} = \theta^\dagger \tau^3 = (d^\ast, u^\ast) \eqno(18) \]
and
\[ \theta = C\overline{\theta}^T = \tau^1 \left( \begin{array}{c} d^\ast \\ u^\ast \end{array} \right) = 
\left( \begin{array}{c}  i u^\ast \\ -i d^\ast \end{array} \right) \eqno(19) \]
and so
\[ u = iu^\ast \eqno(20) \]
\[ \hspace{.3cm}d = -id^\ast \; . \eqno(21) \]
From eq. (15a) we find that 
\[ \Pi^1 = p^1 + x_{,1}^1 - u_{,1} u - d_{,1} d \eqno(22a) \]
\[ \hspace{.2cm}\Pi^2 = p^2 + x_{,1}^2 - id_{,1} u - iu_{,1} d \eqno(22b) \]
\[ \Pi^3 = p^3 + x_{,1}^3 + u_{,1} u - d_{,1} d \eqno(22c) \]
and so we end up with the $N = 1$, $d = 2 + 1$ analogue of eq. (5a)
\[ \chi_u = \pi_u + \Pi^1u + i\Pi^2d + \Pi^3u - 3d_{,1}du \eqno(23a) \]
\[ \chi_d = \pi_d + \Pi^1d + i\Pi^2u - \Pi^3d - 3u_{,1}ud\;. \eqno(23b) \]
These constraints can now be used to form the matrix of PBs
\[
\underset{\sim}{M} = \left( \begin{array}{cc}
\left\lbrace \chi_u, \chi_u\right\rbrace & \left\lbrace \chi_u, \chi_d\right\rbrace  \\
\left\lbrace \chi_d, \chi_u\right\rbrace & \left\lbrace \chi_d, \chi_d\right\rbrace \end{array} \right) = 
\left( \begin{array}{ll}
-2\left(\Pi^1 + \Pi^3 - 4d_{,1}d\right) & -2\left(i\Pi^2 + 2u_{,1} d -  2ud_{,1}\right) \\
 -2\left(i\Pi^2 + 2u_{,1} d -  2ud_{,1}\right) & -2\left(\Pi^1 - \Pi^3 - 4u_{,1}u\right)\end{array} \right) \eqno(24) \]
whose determinant is
\[ \hspace{-3cm}\det \, \underset{\sim}{M} = - 4 \bigg[ \left( \Pi^{1{^2}} + \Pi^{2{^2}} - \Pi^{3{^2}}  \right) - 4\Pi^1 \left( u_{,1} u + d_{,1} d \right) \eqno(25a)\]
\[ - 4i\Pi^2 \left( u_{,1} d - ud_{,1}  \right) - 4\Pi^3
\left( u_{,1} u - d_{,1} d \right)+ 24  u_{,1} u  d_{,1} d \bigg] \]
which, by eqs. (8, 22) becomes
\[ = -8 \left( \Sigma_s + \Sigma_p\right)\;. \eqno(25b) \]

Eliminating both $\chi_u$ and $\chi_d$ is thus not feasible as the DB that would be used is singular when $\Sigma_s = \Sigma_p = 0$. We thus choose to eliminate just $\chi_u$ by  defining the DB 
\[ \left\lbrace A,B \right\rbrace^\ast = \left\lbrace A,B \right\rbrace - \left\lbrace A, \chi_u \right\rbrace \frac{1}{-2(\Pi^1 + \Pi^3 - 4d_{,1}d)}
\left\lbrace \chi_u , B \right\rbrace \;.\eqno(26a) \]
If one had chosen to eliminate $\chi_d$, the DB would be
\[ \left\lbrace A,B \right\rbrace^\# = \left\lbrace A,B \right\rbrace - \left\lbrace A, \chi_d \right\rbrace \frac{1}{-2(\Pi^1 - \Pi^3 - 4u_{,1}u})
\left\lbrace \chi_d , B \right\rbrace \;.\eqno(26b) \]
With the DB of eq. (26a), $\chi_d$ becomes first class, as now by using eq. (24) we see that 
\[  \left\lbrace \chi_d, \chi_d \right\rbrace^\ast  = \det \, \underset{\sim}{M} / \left[ -2 \left(\Pi^1 - \Pi^3 - 4d_{,1}d \right)\right] \approx 0. \eqno(27a) \]
Similarly, from eq. (26b) we find that
\[ \left\lbrace \chi_u , \chi_u \right\rbrace^\# \approx 0.\eqno(27b)\]  The DB of eq. (26a) results in a number of peculiar DBs; for example we find that 
\[  \left\lbrace u, u \right\rbrace^\ast = \frac{1}{2(\Pi^1 + \Pi^3 - 4d_{,1}d)} \eqno(28a) \]
and 
\[  \left\lbrace x_1, x_2 \right\rbrace^\ast = \frac{-iud}{2(\Pi^1 + \Pi^3 - 4d_{,1}d)} \;. \eqno(28b) \]
The lack of commutivity between $x_1$ and $x_2$ in eq. (28b) indicates that we have some of the features of a non-commutative theory (25); we note that the right side of eq. (28b) is field dependent which distinguishes this Dirac Bracket from what is normally encountered in a non-commutative theory.

We now turn to $3 + 1$ dimensions, again in the $N = 1$ limit (ie, we set $\theta^2 = 0$ in eq. (1)).  In this case when $\theta = C\overline{\theta}^T$, we combine eq. (11a) with eq. (A.65) to obtain
\[ \left\lbrace \chi , \overline{\chi} \right\rbrace = - 2 \left( \Pi^\mu + y^\mu \right) \tilde{\gamma}_\mu \eqno(29) \]
where $\Pi^\mu$ and $y^\mu$ are defined as in eq. (15) with $\tilde{\tau}^i$ being replaced by $\tilde{\gamma}^\mu$.  Eq. (29) is the obvious analogue of eq. (14); again, as in $2 + 1$ dimensions, elimination of all of the components of $\chi$ through use of a DB is not possible when on the constraint surface $\Sigma_s = \Sigma_p = 0$.

As $\chi$ now has four components rather than just two, it is not as straight forward now to split these components into two groups, with one group being classified as being second class constraints and the remaining 
group becoming first class.  In examining the superparticle, this was done by looking at the components explicitly [13].  Here, we will use projection operators that are field-independent.

The projection operators
\[ P_\pm = \frac{1}{2} ( 1 \pm\gamma^5 ) \eqno(30) \]
are an obvious candidate for being the projection operators to effect a grouping of the components of $\chi$ into first and second class constraints.  However, by eq. (29)
\[\hspace{-4.2cm} \left\lbrace P_+ \chi , \left( \overline{P_+ \chi}\right)\right\rbrace  = P_+ \left\lbrace \chi , \overline{\chi} \right\rbrace P_-\nonumber \]
\[= -2 \left( \Pi_\mu + y_\mu \right) P_+ \tilde{\gamma}^\mu P_-\; ; \nonumber \]
by eq. (A.3) this becomes 
\[ = - 2 \left( \Pi_\mu + y_\mu \right) \left( \begin{array}{cc}
\left( \begin{array}{cc} 0 & 0 \\ -i\sigma & 0\end{array} \right), \left( \begin{array}{cc} 0 & 0 \\ i & 0 \end{array} \right)\end{array} \right) .\eqno(31a) \]
Similarly we find that 
\[ \left\lbrace P_- \chi , \left( \overline{P_- \chi}\right)\right\rbrace
 = - 2 \left( \Pi_\mu + y_\mu \right) \left( \begin{array}{cc}
\left( \begin{array}{cc} 0 & i\vec{\sigma} \\ 0 & 0\end{array} \right), \left( \begin{array}{cc} 0 & i \\ 0 & 0 \end{array} \right)\end{array} \right) .\eqno(31b) \]
while
\[ \left\lbrace P_+ \chi , \left( \overline{P_- \chi}\right)\right\rbrace
= \left\lbrace P_- \chi , \left( \overline{P_+ \chi}\right)\right\rbrace
= 0. \eqno(31c) \]
By eq. (31) we see that if we eliminate $P_+\chi \;\,(P_-\chi)$ by use of a DB then $P_-\chi \;\,(P_+\chi)$ does not become first class. Thus the projection operators of eq. (30) are not suitable for distinguishing between the first and second class constraints that reside in $\chi$.

Another projection operator that can be used is 
\[ \tilde{P}_\pm = \frac{1}{2}\left( 1 \pm i \gamma^3\gamma^4 \right). \eqno(32) \]
This pair of projection operators satisfies the equations 
\[ \tilde{P}_+ \tilde{\gamma}^\mu\tilde{P}_+ = \tilde{P}_- \tilde{\gamma}^\mu \tilde{P}_- = \left( \tilde{\gamma}^1, \tilde{\gamma}^2, 0, 0 \right) P_\pm \eqno(33a) \]
\[ \tilde{P}_- \tilde{\gamma}^\mu\tilde{P}_+ = \tilde{P}_+ \tilde{\gamma}^\mu \tilde{P}_- = \left( 0, 0 , \tilde{\gamma}^3, \tilde{\gamma}^4 \right) P_\pm \eqno(33b) \]
and so
\[\hspace{-5.7cm} \left\lbrace \tilde{P}_+ \chi , \left( \overline{\tilde{P}_+ \chi}\right)\right\rbrace = \tilde{P}_+ \left[ -2 \left( \Pi_\mu + y_\mu \right) \tilde{\gamma}^\mu \right) \tilde{P}_- \nonumber \]
\[ = \Lambda_\mu \cdot \left( 0, 0, \tilde{\gamma}^3, \tilde{\gamma}^4\right) P_-\quad \left( \Lambda_\mu \equiv -2
\left( \Pi_\mu + y_\mu \right)\right) \eqno(34a) \] 
\[\hspace{-6.8cm} \left\lbrace \tilde{P}_- \chi , \left( \overline{\tilde{P}_- \chi}\right)\right\rbrace = \Lambda_\mu \left( 0, 0, \tilde{\gamma}^3, \tilde{\gamma}^4\right)P_+ \eqno(34b) \]
as well as 
\[\hspace{-3.8cm} \left\lbrace \tilde{P}_+ \chi , \left( \overline{\tilde{P}_- \chi}\right)\right\rbrace = \left\lbrace \tilde{P}_- \chi , \left( \overline{\tilde{P}_+ \chi}\right)\right\rbrace \nonumber \]
\[ =  \Lambda_\mu \cdot \left(  \tilde{\gamma}^1, \tilde{\gamma}^2, 0, 0,\right)P_\pm \; .\eqno(34c) \]
From eq. (34) we find that upon eliminating the constraint $\tilde{P}_+\chi$ by use of the DB 
\[ \left\lbrace A,B \right\rbrace^\ast = \left\lbrace A,B \right\rbrace - 
\left\lbrace A, \overline{\chi} \tilde{P}_- \right\rbrace \frac{1}{\Lambda_3 \tilde{\gamma}^3 + \Lambda_4 \tilde{\gamma}^4} \left\lbrace 
\tilde{P}_+ \chi , B \right\rbrace \eqno(35) \]
then with this DB, $\tilde{P}_-\chi$ becomes first class.  (If $\tilde{P}_- \chi$ were used to define a DB, then $\tilde{P}_+\chi$, with this DB, would be first class.)  It is apparent that the classification of the primary Fermionic constraints can be done in the same way in $2 + 1$ and $3 + 1$ dimensions.  We now turn to $9 + 1$ dimensions.

In $9 + 1$ dimensions, $\theta (\sigma, \tau)$ is both Majorana and Weyl as in eq. (A.67); we can therefore replace $\theta$ by $P_+\theta \equiv \frac{1}{2} (1 + B^{11})\theta$.  As a result, we find that the constraint $\chi^I$ of eq. (5) in $9 + 1$ dimensions can be replaced by 
$P_- \chi^I \equiv \frac{1}{2} (1 - B^{11})\chi^I$ (and $\overline{\chi}^I$ by $\overline{\chi}^IP_+$).  This means that the $N = 1$ limit of eq. (11a) can be expressed in the form 
\[\hspace{-4cm} \left\lbrace \chi , \overline{\chi} \right\rbrace = P_- \left\lbrace \chi , \overline{\chi} \right\rbrace P_+ \nonumber \]
\[ = -2 \left(p^A + x_{,1}^A + \overline{\theta}_{,1} \tilde{B}^A\theta \right) P_- \tilde{B}_A P_+ \nonumber \]
\[ + 2 P_- \left[ \tilde{B}^A\theta_{,1} \overline{\theta} \tilde{B}_A - 
\tilde{B}^A\theta , \overline{\theta}_{,1} \tilde{B}_A  \right]P_+ \nonumber \]
which the Fierz identity of eq. (A.68) reduces to 
\[ = -2 \left(p^A + x_{,1}^A + 2 \overline{\theta}_{,1} \tilde{B}^A\theta \right)\tilde{B}_A P_+\;. \eqno(36) \]
As in $2 + 1$ and $3 + 1$ dimensions, a DB cannot be defined using eq. (36) on the constraint surface $\Sigma_s = \Sigma_p = 0$.  We can however make use of projection operators 
\[ \tilde{P}_\pm = \frac{1}{2} \left( 1 \pm iB^9 B^{10} \right) \eqno(37) \]
which are the $9 + 1$ dimensional analogues of the $3 + 1$ dimensional projection operators  defined in eq. (32).  Since both $P_\pm$ commute with both $\tilde{P}_\pm$, it is possible to consider the PBs
\[\hspace{-8cm}\left\lbrace \tilde{P}_+ \chi , \left( \overline{\tilde{P}_+ \chi}\right)\right\rbrace =   \tilde{P}_+ \left\lbrace \chi , \overline{\chi} \right\rbrace \tilde{P}_- \nonumber \]
\[ \hspace{4cm}= -2\left(p^A + x_{,1}^A + 2 \overline{\theta}_{,1} \tilde{B}^A \theta \right)\left( \left( 0,0,0,0,0,0,0,0, \tilde{B}_9, \tilde{B}_{10}\right)\right)P_+ \tilde{P}_- \nonumber \]
\[\hspace{-4cm}= \left\lbrace \tilde{P}_- \chi , \left( \overline{\tilde{P}_- \chi}\right)\right\rbrace \eqno(38a) \]
while
\[\hspace{-4.7cm}\left\lbrace \tilde{P}_+ \chi , \left( \overline{\tilde{P}_- \chi}\right)\right\rbrace = \left\lbrace \tilde{P}_- \chi , \left( \overline{\tilde{P}_+ \chi}\right)\right\rbrace   \nonumber \]
\[ = -2\left(p^A + x_{,1}^A + 2 \overline{\theta}_{,1} \tilde{B}^A \theta \right)\nonumber \]
\[\hspace{4cm}\left( \left( \tilde{B}_1, \tilde{B}_2,\tilde{B}_3, \tilde{B}_4,\tilde{B}_5, \tilde{B}_6,\tilde{B}_7, \tilde{B}_8, 0, 0 \right)\right)P_+\tilde{P}_\pm\;. \eqno(38b) \]
Just as in $3 + 1$ dimensions, we can now define a DB to eliminate $\tilde{P}_+\chi$; with this DB $\tilde{P}_-\chi$ becomes a first class constraint.  The roles of $\tilde{P}_+\chi$ and $\tilde{P}_-\chi$ can be reversed.

Extending these considerations from the $N = 1$ to the $N = 2$ superstring
is straightforward.  By the Fierz identity of eq. (A.68), eqs. (11a,b) become 
\[ \left\lbrace \chi^1, \overline{\chi}^1 \right\rbrace = -2 \left( p^A + x_{,1}^A + 2 \overline{\theta}_{,1}^1 \tilde{B}^A \theta^1 \right) \tilde{B}_AP_+ \nonumber \]
\[\hspace{-.35cm} = -2 \left( \Pi^A + y^A \right) \tilde{B}_A P_+ \eqno(39a) \]
\[ \left\lbrace \chi^2, \overline{\chi}^2 \right\rbrace = -2 \left( p^A - x_{,1}^A - 2 \overline{\theta}_{,1}^2 \tilde{B}^A \theta^2 \right) \tilde{B}_AP_+ \nonumber \]
\[\hspace{-.35cm} = -2 \left( \Pi^A - y^A \right) \tilde{B}_A P_+\; .
\eqno(39b) \]
We see that $\tilde{P}_\pm \chi^I$ can now be considered as being second class; the resulting DB results in $\tilde{P}_\mp \chi^I$ being first class.

The first class constraints present in a theory can be used to find the generator of a gauge transformation that leaves the action invariant. There are several approaches to finding this generator [9, 10]; we will examine the latter (the ``HTZ'' approach)for the $N = 1$ superstring  in $ 2 + 1$ dimensions.

The form of the generator is 
\[ G = \int \left( a I\!\!P + a^1 I\!\!P_1 + \overline{b} \chi_d + A_s \Sigma_s + A_p \Sigma_p \right) d\sigma \eqno(40)\]
if we treat $\chi_u$ in eq. (23a) as being second class.  With the canonical Hamiltonian of eq. (6), we see that the total Hamiltonian is given by 
\[ H_T = \int \left( \mathcal{H}_C + U I\!\!P + U^1 I\!\!P_1 +\overline{V}_d \chi_d \right) d\sigma \; ,\eqno(41) \]
where $U$, $U^1$ and $V_d$ are Lagrange multipliers. The HTZ equation that fixes the gauge functions $a$, $a^1$, $b$, $A_s$ and $A_p$ in $G$ is [10]
\[ \frac{Da}{Dt} I\!\!P + \frac{Da^1}{Dt} I\!\!P +  \frac{D\overline{b}}{Dt} \chi_d +  \frac{DA_s}{Dt} \Sigma_s +  \frac{DA_p}{Dt} \Sigma_p\nonumber \]
\[ + \left\lbrace G, H_T \right\rbrace^\ast - \delta a I\!\!P - \delta a^1 I\!\!P_1 - \delta \overline{b} \chi_d = 0 \; .\eqno(42) \]
By using eqs. (12, 13, 27a) we find that eq. (42) can be satisfied if 
\[ a = \dot{A}_s - \frac{4\overline{b}\,\overline{V}_d}{\Pi^1 - \Pi^3 - 4d_{,1}d} + \left( A_p H_{,1} - A_{p,1}H\right) + \left( A_s H_{,1}^1 - A_{s,1} H^1\right) \eqno(43a)\]
\[ a^1 = \dot{A}_p - \frac{4\overline{b}\,\overline{V}_d}{\Pi^1 - \Pi^3 - 4d_{,1}d} + \left( A_p H_{,1}^1 - A_{p,1}H^1\right) + \left( A_s H_{,1} - A_{s,1} H\right) \eqno(43b)\]
\[ \delta U = \dot{a}\eqno(44a)\]
\[ \delta U^1 = \dot{a}^1\eqno(44b)\]
\[ \delta \overline{V}_d = \dot{\overline{b}}\; .\eqno(44c)\]
An analogous set of equations arises when finding the gauge generator for the superparticle [13].  As with the superparticle, a second set of gauge transformations can be found by reversing the roles of $\chi_u$ and $\chi_d$; that is we use the DB of eq. (26b) rather than eq. (26a).  The $\overline{b}$ dependent parts of these two sets of gauge transformations are presumably related to one half of the $N = 1$, $2 + 1$ dimensional version of the Fermionic $\kappa$-symmetry present in the $9 + 1$ dimensional superstring. This is what happens with the superparticle.

By use of eq. (43), $G$ in eq. (40) is fixed in terms of the gauge functions $\overline{b}$, $A_s$, $A_p$ as well as the Lagrange multiplier $\overline{V}_d$.  As with the superparticle, one can fix $\overline{V}_d$ by ensuring that the transformation $\delta A = \left\lbrace A, G. \right\rbrace^\ast$ made on each dynamical variable $A$ leaves the action invariant.

The set of gauge transformations resulting from the first class constraints present in the $2 + 1$ and $9 + 1$ dimensional superstring can similarly be found.

We now turn to examining the superstring with other space-time dimensions for the target space with the aim of finding a simpler Fermionic gauge transformation.

\section{The Simpler Superstring}

In order to simplify the superstring action of eq. (1), we will examine when it becomes possible for
\[ \overline{\lambda}\tilde{\rho}^\mu \chi = \overline{\chi} \tilde{\rho}^\mu \lambda \eqno(45) \]
when $\lambda$ and $\chi$ are Majorana spinors. Eqs. (45) leads to a simplification of eq. (1), as 
\[\hspace{-2cm} \overline{\theta}_{,a} \tilde{\rho}^\mu \theta = \overline{\theta}\tilde{\rho}^\mu \theta_{,a} \nonumber \]
\[ = \frac{1}{2} \left( \overline{\theta} \tilde{\rho}^\mu \theta \right)_{,a} \eqno(46)\]
and so, upon discarding a surface term, the action of eq. (1) in the $N = 1$ limit  becomes when the target space has $d$ dimensions
\[ S = \int d\tau d\sigma \left[ \frac{1}{2} h^{ab} \left( x^\mu + \frac{1}{2}\overline{\theta}\tilde{\rho}^\mu\theta \right)_{,a} \left(x_\mu + \frac{1}{2} \overline{\theta}\tilde{\rho}_\mu \theta \right)_{,b} \right]\eqno(47) \]
(since $\epsilon^{ab} A_{,a} B_{,b} = \epsilon^{ab} (AB_{,b})_{,a} - \epsilon^{ab} AB_{,ab}$).
The obvious local Fermionic gauge invariance possessed by this action is
\[\hspace{-1.3cm}\theta \rightarrow \theta + \epsilon \eqno(48a) \]
\[ x^\mu \rightarrow x^\mu - \overline{\epsilon}\tilde{\rho}^\mu \theta - \frac{1}{2} \overline{\epsilon} \,\tilde{\rho}^\mu\epsilon\; . \eqno(48b) \]
showing that the Fermionic field $\theta$ is purely gauge when eq. (45) holds, reducing the model to being purely Bosonic.

We now will determine when eq. (45) can hold. If $E$ denotes either $C$ or $D$ in eqs. (A.18, 19) then
\[ \hspace{-5.8cm}\overline{\lambda} \tilde{\rho}^\mu \chi = \left(\lambda^T E^{-1T}\right) \left[ (-1)^\epsilon E \tilde{\rho}^{\mu T} E^{-1} \right] E \overline{\chi}^T  \eqno(49) \]
\[= (-1)^{\epsilon + 1} \overline{\chi} \tilde{\rho}^\mu (E^T E^{-1})^T \lambda \quad (\epsilon = 1 \;\; \rm{for} \;\; E = C,\; 
 \epsilon = 0\;\;\rm{for} \;\; E = D).\nonumber \]
From eq. (49), eq. (45) holds provided
\[ E = (-1)^{\epsilon + 1} E^T \\;.\eqno(50) \]
From eqs. (A.21, A.22, A.25, A.26, A.31, A.32, A.39, A.40) it follows that the only candidates are $E = D$ in $4d$, $E = C$ or $D$ in $6d$, $E = C$ in $8d$, with neither $C$ nor $D$ in $10d$.

We need to ensure that spinors can be Majorana; for this condition to be consistent we must have
\[\hspace{-3.3cm} \theta = (\theta_E)_E \eqno(51) \]
which means that
\[\hspace{-1.3cm} = \left[ E\left(\theta^\dagger A^{-1}\right)^T\right]_E \nonumber \]
\[ = E\left\lbrace \left[ E A^{-1T}\theta^\ast \right]^\dagger A^{-1}\right\rbrace^T \nonumber \]
\[\hspace{-1.1cm} = E\, A^{-1T} E^\ast A^{-1\dagger}\theta \;.\]
Consequently, for a spinor to be Majorana, we need to have
\[ EA^{-1T} E^\ast A^{-1\dagger} = 1 \eqno(52) \]
From eqs. (A.20-A.48) and eq. (52) we see that the Majorana condition is satisfied if $\theta = \theta_C$ in $3 + 3$, $5 + 3$ or $4 + 4$ dimensions or if $\theta = \theta_D$ in $4 + 2$ or $3 + 3$ dimensions.

A further restriction that can be placed on a spinor $\theta$ is that it be Weyl in addition to being Majorana.  Consequently, in $d$ dimensions $\rho^{d+1}\theta = \theta$ so that  $\rho^{d+1}\theta_E = \theta_E$.  As a result we have 
\[ \rho^{d+1}(E \overline{\theta}^T) = \left[ EA^{-1T}, \rho^{d+1} \right]\theta^{i\ast} + E (\rho^{d+1}\theta)^\dagger A^{-1} \eqno(53) \]
and so if $\theta$ is to be both Majorana and Weyl,
\[ \left[ E A^{-1T}, \rho^{d+1} \right] = 0 \;.\eqno(54)\]
From the properties of $A$, $C$, $D$ given in appendix A, this additional restriction limits our attention to $\theta$ satisfying $\theta = \theta_C$ in $3 + 3$ or $4 + 4$ dimensions, or $\theta = \theta_D$ in $3 + 3$ dimensions. However, in $4 + 4$ dimensions, if $\lambda$ and $\chi$ are both Majorana and Weyl, then $\overline{\lambda} \tilde{\beta}^\alpha\chi$ vanishes.

We are left with $\theta$ being Weyl with either $\theta = \theta_C$ or 
$\theta = \theta_D$ in $3 + 3$ dimensions.  (In this case, $\overline{\theta}_{,1}\tilde{\Gamma}^M\theta$ is Hermitian.)  Since, if $\theta^I = \theta^I_E = E \overline{\theta}^{IT}$, it follows that
$\overline{\theta}^{IT} = E^{-1}\theta^I$, $\overline{\theta}^{I} = \theta^{IT}E^{-IT}$ and $\theta^{IT} = \overline{\theta}^{I}  E^T$, and hence 
\[\hspace{-1.1cm} \left\lbrace \pi^I, \overline{\theta}^J\right\rbrace = - \delta^{IJ} \eqno(55a) \]
leads to 
\[\hspace{-.7cm} \left\lbrace \pi^I,\theta^{JT}\right\rbrace = - \delta^{IJ}E^T \eqno(55b) \]
\[\hspace{.3cm} \left\lbrace \theta^I, \overline{\pi}^J\right\rbrace = - \delta^{IJ}E E^{-1T} \eqno(55c) \]
\[ \left\lbrace \overline{\theta}^{IT}, \overline{\pi}^J\right\rbrace = - \delta^{IJ}E^{-1T}\; . \eqno(55d) \]
With
\[\hspace{-1.4cm} \chi^I = \pi^I - \left[ p^\mu - (-1)^I \left( x_{,1}^\mu + \overline{\theta}_{,1}^I \tilde{\rho}^\mu \theta^I\right)\right]\tilde{\rho}_\mu \theta^I \eqno(56a) \]
\[ \overline{\chi}^I = \overline{\pi}^I + (-1)^{t+1} \overline{\theta}^I \tilde{\rho}_\mu \left[ p^\mu - (-1)^I \left( x_{,1}^\mu + \overline{\theta}_{,1}^I \tilde{\rho}^\mu \theta^I\right)\right] \eqno(56b) \]
we find that in $d = s + t$ dimensions
\[ \left\lbrace \chi^1, \overline{\chi}^2 \right\rbrace = 0 \eqno(57a) \]
and 
\[\hspace{-4.3cm} \int d\sigma d\sigma^\prime f(\sigma) \left\lbrace \chi^1 (\sigma), \overline{\chi}^1 (\sigma^\prime) \right\rbrace g(\sigma^\prime) \nonumber \]
\[= \int d\sigma \bigg[ fg \left(p_\mu + x_{\mu ,1} + \overline{\theta}_{,1}^1 \tilde{\rho}^\mu \theta^1 \right) \tilde{\rho}^\mu \left(EE^{-1T} + (-1)^t\right)\]
\[ + (-1)^t \left( f Z_\mu (\overline{Z}^\mu g)_{,1} - (fZ_\mu)_{,1} \overline{Z}^\mu g\right) \]
\[ + f Z_\mu \overline{Z}_{,1}^\mu EE^{-1T} g + (fZ_\mu Z^{\mu T})_{,1} E^{-1T} g \]
\[ + (-1)^{t+1} \left(f \left(Z_\mu \overline{Z}^\mu g\right)_{,1} + 
f E^T \overline{Z}_{\mu ,1}^T \overline{Z}^\mu g\right) \bigg].\eqno(57b)\]
In eq. (57b), $f$ and $g$ are test functions (used as in ref. [18]) and $Z^\mu = \tilde{\rho}^\mu \theta^1$, $\overline{Z}^\mu = \overline{\theta}^1\tilde{\rho}^\mu $.

If $\theta^1 = \theta^1_C$ and is Weyl in $3 + 3$ dimensions, then by eq. (A.25) 
\[\left\lbrace \chi^1 , \overline{\chi}^1 \right\rbrace = 0.\eqno(58) \]
However, if $\theta = \theta_D$ and is Weyl in $3 + 3$ dimensions, then by eq. (A.26)
\[\hspace{-4.3cm} \int d\sigma d\sigma^\prime f(\sigma) \left\lbrace \chi^1 (\sigma), \overline{\chi}^1 (\sigma^\prime) \right\rbrace g(\sigma^\prime) \nonumber \]
\[= \int d\sigma \bigg[ -2 \left(p^M + x_{,1}^M + \overline{\theta}_{,1}^1 \tilde{\Gamma}^M \theta^1 \right) \tilde{\Gamma}_M fg   \eqno(59)\]
\[ + \left( fZ^M\right)_{,1}  \left( \overline{Z}_M g\right) -
\left(f Z^M \right) \left( \overline{Z}_M g\right)_{,1} \bigg].\]
(Eqs. (58, 59) can be seen more immediately using the action of eq. (47) directly.)

Eq. (59) can be simplified using the Fierz identities that follow from eqs. (A.69-A.77).  We find that 
\[\hspace{-4cm}(f Z^M)_{,1} (\overline{Z}_Mg) - (f Z^M) (\overline{Z}_Mg)_{,1} \]
\[ = fg \left( Z^M_{,1} \overline{Z}_M - Z^M \overline{Z}_{M,1} \right) + 
Z^M \overline{Z}_M \left( f_{,1} g - f g_{,1} \right)\]
\[ = - \left( \overline{\theta} \tilde{\Gamma}^M \theta \tilde{\Gamma}_M  - \tilde{\theta} \tilde{\Gamma}^M \Gamma^7 \theta \tilde{\Gamma}_M \Gamma^7 \right) (f_{,1}g - fg_{,1}) \]
which by the Weyl property of $\theta$ becomes
\[= - \overline{\theta} \tilde{\Gamma}^M \theta \tilde{\Gamma}_M (1 - \Gamma^7 ) (f_{,1}g - fg_{,1}). \eqno(60) \]
However, as $\overline{\chi} \left[\frac{1}{2} (1 - \Gamma^7)\right] = 0$ we see that eq. (60) results in eq. (59) to simply reducing to 
\[\hspace{-3cm} \int d\sigma d\sigma^\prime P_- f(\sigma) \left\lbrace \chi^1 (\sigma), \overline{\chi}^1(\sigma^\prime) \right\rbrace g(\sigma^\prime) P_+ \eqno(61) \]
\[ = \int d\sigma \left[ -2 \left( p^M + x_{,1}^M + \overline{\theta}_{,1}^1 \tilde{\Gamma}^M \theta^1 \right) \tilde{\Gamma}_M fg\right] P_+ \]
\[ \hspace{3cm} \left( P_\pm \equiv \frac{1}{2} (1 \pm \Gamma^7)\right).\]

With eq. (61), all of the constraints $\chi^I$ are seen to be second class when $\theta = \theta_D$ and is Weyl in $3 + 3d$.  We note the difference in the canonical structure of the superstring action in $3 + 3$ dimensions when $\theta = \theta_C$ and Weyl, and when $\theta = \theta_D$ and Weyl. In the former case, the primary Fermionic constraints are all first class and their contribution to the gauge generator leads to the gauge transformation of eq. (48).  In the latter case, the primary Fermionic constraints are second class and the gauge invariance of eq. (48) is not associated with any first class constraint.  (We note that with the Palatini action in $1 + 1$ dimensions, the first class constraints are associated with an unusual gauge transformation while the gauge transformation associated with the diffeomorphism invariance of the action is not associated with any first class constraints [20].)

Finally we shall demonstrate how in five dimensions, no first class Fermionic constraints arise and consequently, there is an absence of Fermionic gauge symmetries.  We use matrices $g^m = (\gamma^\mu , \gamma^5)$ defined in $5 + 0$ dimensions by eq. (A.3).  The matrix $C$ of eq. (A.18) does not exist in $5d$, while $D$ of eq. (A.19) is given by
\[ D = g^1g^3 \,.\eqno(62) \]
In $4 + 1d$ where $A$ of eq. (A.12) is
\[ A = g^5 \eqno(63) \]
we find that if $\theta = \theta_D = g^1g^3g^5\theta^\ast$, then $(\theta_D)_D = -\theta$ and hence $\theta = 0$; consequently we will not consider $4 + 1d$ any further as now $\theta$ and $\overline{\theta}$ are independent.

For $3 + 2d$ where $A = g^4g^5$, $\theta = \theta_D = g^2\theta^\ast$, it is feasible to set $\theta = \theta_D$.  However, we now have 
\[ \left( \overline{\lambda} \tilde{g}^m\chi \right)^\dagger = - \overline{\chi} \tilde{g}^m\lambda \eqno(64) \]
if $\lambda = \lambda_D$, $\chi = \chi_D$ and hence in order to have an Hermitian action, the $N = 1$ version of eq. (1) must become
\[ S = \int d\tau\,d\sigma \left[ \frac{1}{2} h^{ab} Z_a^m Z_{mb} - i \epsilon^{ab} x_{,a}^m \overline{\theta}_{,b}\, \tilde{g}_m\theta \right] \eqno(65) \]
where now
\[Z^m_a = x_{,a}^m + i\overline{\theta}_{,a} \,\tilde{g}^m\theta \,.\eqno(66) \]
If $\pi = \frac{\partial \mathcal{L}}{\partial \overline{\theta}_{,0}}$ then much like eq. (4c) we find that we have the primary Fermionic constraint
\[ \chi = \pi - i(p^m + z^m)\tilde{g}_m\theta \eqno(67) \]
where $z^M = Z_1^M$.  Since $\theta = \theta_D$, it follows from eq. (67) that $\pi = -\pi_D$; consequently as
\[ \left\lbrace \pi , \overline{\theta}\right\rbrace = -1\eqno(68a) \]
from eq. (B.7), it follows that
\[ \left\lbrace \pi , \overline{\theta}^T\right\rbrace = D, \qquad
\left\lbrace \theta , \overline{\pi}\right\rbrace = -1, \qquad
\left\lbrace  \overline{\theta}^T, \overline{\pi}\right\rbrace = D\,.\eqno(68b,c,d) \]
Eqs. (67, 68) now lead to 
\[ \left\lbrace \chi , \overline{\chi}\right\rbrace = -2i (p^m + z^m)\tilde{g}_m + 2 \left( \tilde{g}^m \theta \overline{\theta}_{,1} \,\tilde{g}_m - \tilde{g}^m \theta_{,1} \overline{\theta} \,\tilde{g}_m
\right)\,; \eqno(69) \]
the Fierz identity
\[ (\tilde{g}^m)_{ij} (\tilde{g}_m)_{k\ell} = \delta_{i\ell} \delta_{kj} - \frac{3}{4}  (\tilde{g}^m)_{i\ell} (\tilde{g}_m)_{kj} - \frac{1}{8} 
(\tilde{g}^{[m}\tilde{g}^{n]})_{i\ell}
(\tilde{g}_{[m}\tilde{g}_{n]})_{kj} \eqno(70) \]
then reduces eq. (69) to
\[ \left\lbrace \chi , \overline{\chi}\right\rbrace = -2i (p^m + z^m)\tilde{g}_m + \frac{1}{2•} \overline{\theta}_{,1} 
\tilde{g}^{[m}\tilde{g}^{n]} \,\theta\,
\tilde{g}_{[m}\tilde{g}_{n]} \eqno(71) \]
since $\theta = \theta_D$.  From eq. (71) we see that unlike what happens with eqs. (14, 29) a DB can be defined from this PB even when the secondary Bosonic constraints are satisfied.  Consequently all primary Fermionic constraints are second class in $3 + 2d$ and no Fermionic gauge symmetry exists. However, the action of eq. (1) when $N = 1$ and $\theta = \theta_D$ in $3 + 2$ dimensions is globally supersymmetric under the transformation $x^m \rightarrow x^m + i\overline{\epsilon} \tilde{g}^m\theta$, $\theta \rightarrow \theta - \epsilon (\epsilon = \epsilon_D)$ provided surface terms are discarded.  This can be established quite easily once we note that eqs. (45) and (64) both hold in $3 + 2$ dimensions when all spinors satisfy $\theta = \theta_D$, as then $\epsilon^{ab} \overline{\theta}_{,a} \tilde{g}^m \theta_{,b} = 0$.

\section{Discussion}

In this paper we have examined the canonical structure of the superstring action.  If the target space is $2 + 1$ or $3 + 1$ dimensional and the spinors are Majorana, and when the target space is $9 + 1$ dimensional and the spinors are Majorana-Weyl, then the primary Fermionic constraints can be divided into two groups using field-independent projection operators with one group being second class, the other first class, with the first class constraints generating a gauge transformation.  We also find that with a $3 + 3$ dimensional target space, and the spinors $\theta$ are Majorana-Weyl, the Fermionic gauge transformation greatly simplifies and that the canonical structure when $\theta = \theta_C$ and $\theta = \theta_D$ are different. In $5d$, no Fermionic first class constraints occur and no Fermionic gauge symmetry exists.

When performing a canonical analysis of any theory, manifest covariance is necessarily lost as the ``time'' coordinate is given special status.  In our treatment of 
superstring theory, this problem is exacerbated by having been forced to distinguish between different components of the primary spinorial constraints, further obscuring the manifest covariance in the target space that is manifest in the initial action of eq. (1).  However, the lack of manifest covariance in the canonical formalism doesn't preclude the dynamics from being covariant.

To address this problem, we follow Dirac [23] and examine the generators of the Poincar$\acute{\rm{e}}$ group in the target space.  In $3 + 1$ dimensions, the generators are given by [15]
\[M_{\mu\nu} = L_{\mu\nu} + S_{\mu\nu} \eqno(72)\]
where
\[ L_{\mu\nu} = \int \left( x_\mu p_\nu - x_\nu p_\mu \right)d\sigma \eqno(73a) \]
\[ S_{\mu\nu} = i \int \left( \overline\theta \Sigma_{\mu\nu} \pi \right) d\sigma \quad \left( \Sigma_{\mu\nu} \equiv \frac{i}{4} \left[ \gamma_\mu , \gamma_\nu \right] \right) \eqno(73b) \]
and
\[ P_\mu = \int p_\mu d\sigma \; . \eqno(74) \]
These satisfy the PB algebra
\[ \left\lbrace M_{\mu\nu}, M_{\lambda\sigma} \right\rbrace = \eta_{\mu\lambda} M_{\nu\sigma} - \eta_{\nu\lambda} M_{\mu\sigma} + \eta_{\nu\sigma} M_{\mu\lambda} - \eta_{\mu\sigma} M_{\nu\lambda} \eqno(75a) \]
\[\hspace{-3cm} \left\lbrace M_{\mu\nu}, P_\lambda \right\rbrace = \eta_{\mu\lambda} P_\nu - \eta_{\nu\lambda} P_\mu \eqno(75b) \]
\[\hspace{-4cm} \left\lbrace P_\mu , P_\nu \right\rbrace = 0\; .\eqno(75c) \]
Since the constraint $\chi$ is a spinor in target space, we find that
\[ \left\lbrace M_{\mu\nu}, \chi\right\rbrace = i \Sigma_{\mu\nu} \chi \eqno(76a) \]
\[ \left\lbrace \overline{\chi}, M_{\mu\nu} \right\rbrace = i \overline{\chi} \Sigma_{\mu\nu} \chi \eqno(76b) \]
and hence we obtain the DB (from eq. (35))
\[ \hspace{-5cm}\left\lbrace M_{\mu\nu}, M_{\lambda\sigma} \right\rbrace^\ast = 
\left\lbrace M_{\mu\nu}, M_{\lambda\sigma} \right\rbrace \]
\[ + \bigg[ \overline{\chi} \bigg( \Sigma_{\mu\nu} \tilde{P}_- \frac{1}{\Lambda_3 \tilde{\gamma}^3 + \Lambda_4 \tilde{\gamma}^4} \tilde{P}_+ \Sigma_{\lambda\sigma} \]
\[ -\Sigma_{\lambda\sigma} \tilde{P}_-
\frac{1}{\Lambda_3 \tilde{\gamma}^3 + \Lambda_4 \tilde{\gamma}^4}
\tilde{P}_+ \Sigma_{\mu\nu} \bigg)\chi \bigg]\; ,\eqno(77) \]
It is evident that on the constraint surface $\chi = 0$, when both the first class constraints $\tilde{P}_-\chi$ and the second class constraints $\tilde{P}_+\chi$ vanish, the PB and DB for $M_{\mu\nu}$ and $P_\lambda$ coincide.  (The same holds true in $2 + 1$ and $9 + 1$ dimensions.)  

Quantization of the superstring using the canonical structure of its classical action should now be attempted. If the Dirac quantization procedure [7,8] (in which the classical DB becomes the quantum commutator and the first class constraints annihilate physical states) can be carried through unambiguously without the complications arising from operator ordering problems, then Poincar$\acute{\rm{e}}$ invariance on the target space is retained on the constraint surface.  This is currently being considered.

In ref. [23], Dirac has discussed the possibility of quantizing when using ``light-cone'' coordinates as an alternative to the ``space-time'' coordinates that have been employed in this paper.  This option has been used when quantizing the superstring.  However, in this approach to the superstring, the equations of motion have been used to eliminate both Bosonic and Fermionic degrees of freedom prior to applying the canonical formalism in conjunction with light cone coordinates.  The projection operator used to eliminate Fermionic degrees of freedom in the light cone approach (see section 5.2.1 in ref. [1]) is similar to that of eq. (37) which is used to separate first and second class constraints. We note that when one simply eliminates second class constraints by hand prior to applying the canonical formalism, one is unable to form Dirac Brackets, or to find the first class constraints that lead to the generators of Fermionic gauge transformations.  The non-commuting Dirac Brackets of eq. (28) would also be lost. It would be interesting to apply the full Dirac constraint formalism to the superstring when using light-cone coordinates without initially eliminating degrees of freedom to simplify the action.  Such an analysis of Yang-Mills theory has been performed in ref. [26]. 

\section*{Acknowledgements}
Roger Macleod had many helpful suggestions.

\section*{Appendix A}

In this appendix we consider spinors in various space-time dimensions [3, 21].  If we start with the Pauli spin matrices
\[ \sigma^1 = \left( \begin{array}{cc} 0 & 1 \\ 1 & 0 \end{array} \right) \quad 
\sigma^2 = \left( \begin{array}{cc} 0 & -i \\ i & 0 \end{array} \right)
\quad
\sigma^3 = \left( \begin{array}{cc} 1 & 0 \\ 0 & -1 \end{array} \right)
 \eqno(A.1) \]
then in Euclidean space we empty the following conventions for Dirac matrices:
\[\hspace{-3.8cm} 3D:\quad\quad \tau^i = \left( \begin{array}{cc} 0 & i \\ -i & 0 \end{array} \right)\; , \quad 
\left( \begin{array}{cc} -1 & 0 \\ 0 & 1 \end{array} \right)\; ,
\quad
\left( \begin{array}{cc} 0 & 1 \\ 1 & 0 \end{array} \right)
 \eqno(A.2) \]
\[\hspace{-2cm} 4D:\quad\quad \gamma^\mu = \left( \begin{array}{cc} 0 & i\sigma^i \\ -i\sigma^i & 0 \end{array} \right)\; , \quad 
\left( \begin{array}{cc} 0 & 1 \\ 1 & 0 \end{array} \right)\; ;
\quad
\gamma^5 = \gamma^1 \cdots \gamma^4 = \left( \begin{array}{cc} -1 & 0 \\ 0 & 1 \end{array} \right)
 \eqno(A.3) \]
\[ 6D:\quad\quad \Gamma^M = \left( \begin{array}{cc} 0 & i\gamma^\mu \\ -i\gamma^\mu & 0 \end{array} \right)\; , \quad 
\left( \begin{array}{cc} 0 & 1 \\ 1 & 0 \end{array} \right)\; ;
\quad
\Gamma^7 = i\Gamma^1 \cdots \Gamma^6 = \left( \begin{array}{cc} -1 & 0 \\ 0 & 1 \end{array} \right)
 \eqno(A.4) \]
\[ 8D:\quad\quad \beta^\alpha = \left( \begin{array}{cc} 0 & i\Gamma^M \\ -i\Gamma^M & 0 \end{array} \right)\; , \quad 
\left( \begin{array}{cc} 0 & 1 \\ 1 & 0 \end{array} \right)\; ;
\quad
\beta^9 = -\beta^1 \cdots \beta^8 = \left( \begin{array}{cc} -1 & 0 \\ 0 & 1 \end{array} \right)
 \eqno(A.5) \]
\[ 10D:\quad\quad B^A = \left( \begin{array}{cc} 0 & i\beta^a \\ -i\beta^a & 0 \end{array} \right)\; , \quad 
\left( \begin{array}{cc} 0 & 1 \\ 1 & 0 \end{array} \right)\; ;
\quad
B^{11} = -iB^1 \cdots B^{10} = \left( \begin{array}{cc} -1 & 0 \\ 0 & 1 \end{array} \right)\;.
 \eqno(A.6) \]
(In $d$ dimensions, these matrices have $2^{[d/2]}$ rows and columns.)  Taking the Dirac matrix in $d$ dimensions to be $\rho^\mu$, then if
\[\Sigma^{\mu\nu} = \frac{i}{4} \left[ \rho^\mu , \rho^\nu \right] \eqno(A.7) \]
then the transformation
\[ \psi \rightarrow \exp \left( \frac{i}{2} \Sigma^{\mu\nu} \omega_{\mu\nu} \right) \psi \equiv U \psi \eqno(A.8) \]
for the spinor $\psi$ implies that
\[ \psi^\dagger \rightarrow \psi^\dagger \exp \left( -\frac{i}{2} \Sigma^{\mu\nu} \omega_{\mu\nu} \right) =  \psi^\dagger U^{-1} \; .\eqno(A.9) \]
If we move from $d$ dimensional Euclidean space to a space with $t$ time and $s$ space dimensions ($d = s + t$), the metric becomes 
\[ \eta^{\mu\nu} = \delta^{\mu\nu} \;\; \left( 1 \leq (\mu , \nu) \leq s\right) \eqno(A.10) \]
\[ \hspace{1.5cm}= -\delta^{\mu\nu} \left( s + 1 \leq (\mu ,\nu ) \leq d\right) \]
and we replace $\rho^\mu$ with $\tilde{\rho}^\mu$ where 
\[ \tilde{\rho}^\mu = \rho^\mu \;\; (\mu = 1 \cdots s) \eqno(A.11) \]
\[ \hspace{1.8cm}
= i\rho^\mu \;\; (\mu = s + 1 \cdots d) \; . \]
We now define
\[ A = \rho^{s+1} \cdots \rho^d \eqno(A.12) \]
so that
\[ A^{-1} \tilde{\rho}^\mu A = (-1)^t \tilde{\rho}^{\mu^{\dagger}} \; .\eqno(A.13) \]
Consequently, if 
\[ \tilde{\Sigma}^{\mu\nu} = \frac{i}{4} \left[ \tilde{\rho}^\mu , \tilde{\rho}^\nu \right] \eqno(A.14) \]
and
\[ \tilde{U} = \exp \left( \frac{i}{2} \tilde{\Sigma}^{\mu\nu} \omega_{\mu\nu} \right) \eqno(A.15) \]
then
\[ \psi \rightarrow \tilde{U}\psi\; , \quad \overline{\psi} \rightarrow \overline{\psi} \tilde{U}^{-1} \eqno(A.16) \]
where
\[ \overline{\psi} = \psi^\dagger A^{-1} \; . \eqno(A.17) \]
If now $C$ and $D$ are defined so that
\[ C^{-1} \rho^\mu C = - \rho^{\mu T} \eqno(A.18) \]
\[ D^{-1} \rho^\mu D = + \rho^{\mu T} \eqno(A.19) \]
then $\psi_C = C \overline{\psi}^T$ and $\psi_D = D\overline{\psi}^T$ transform as $\psi$ does in eq. (A.16).

More explicitly, we find that 
\[\hspace{-6cm} 4d: \qquad
\gamma^{(1,3)} = -\gamma^{(1,3)T} \quad \gamma^{(2,4)}  = \gamma^{(2,4)T} \eqno(A.20) \]
\[\hspace{-2.6cm}C = \gamma^2\gamma^4 = - C^{-1} = - C^\dagger = -C^T = C^\ast \eqno(A.21) \]
\[\hspace{-2.4cm}D = \gamma^1 \gamma^3 = -D^{-1} = -D^\dagger = -D^T = D^\ast \eqno(A.22) \]
\[\hspace{-2.6cm}\underline{3 + 1} \quad A = \gamma^4 = A^{-1} = A^\dagger = A^T = A^\ast \eqno(A.23)  \]
\[\hspace{-4.6cm} 6d: \qquad
\Gamma^{(2,4,5)} = -\Gamma^{(2,4,5)T}\;,  \quad \Gamma^{(1,3,6)}  = \Gamma^{(1,3,6)T} \eqno(A.24) \]
\[\hspace{-2.4cm}C = \Gamma^2\Gamma^4\Gamma^5 = - C^{-1} = - C^\dagger = C^T = 
-C^\ast \eqno(A.25) \]
\[\hspace{-2.4cm}D = \Gamma^1 \Gamma^3 \Gamma^6 = -D^{-1} = -D^\dagger = -D^T = D^\ast \eqno(A.26) \]
\[\hspace{-2.8cm}\underline{5 + 1} \quad A = \Gamma^6 = A^{-1} = A^\dagger = A^T = A^\ast \eqno(A.27)  \]
\[\hspace{-1.8cm}\underline{4 + 2} \quad A = \Gamma^5\Gamma^6 = A^{-1} = -A^\dagger = A^T = -A^\ast \eqno(A.28)  \]
\[\hspace{-1cm}\underline{3 + 3} \quad A = \Gamma^4\Gamma^5\Gamma^6 = -A^{-1} = -A^\dagger = -A^T = A^\ast \eqno(A.29)  \]
\[\hspace{-4cm} 8d: \qquad
\beta^{(1,3,6,7)} = -\beta^{(1,3,6,7)T}\;,  \quad \beta^{(2,4,5,8)}  = \beta^{(2,4,5,8)T} \eqno(A.30) \]
\[\hspace{-2.7cm}C =  \beta^2\beta^4\beta^5\beta^8 =  C^{-1} =  C^\dagger = C^T = C^\ast \eqno(A.31) \]
\[\hspace{-2.7cm}D = \beta^1\beta^3\beta^6\beta^7  = D^{-1} = D^\dagger = D^T = D^\ast \eqno(A.32) \]
\[\hspace{-2.8cm}\underline{7 + 1} \quad A = \beta^8 = A^{-1} = A^\dagger = A^T = A^\ast \eqno(A.33)  \]
\[\hspace{-1.5cm}\underline{6 + 2} \quad A = \beta^7\beta^8 = -A^{-1} = -A^\dagger = A^T = -A^\ast \eqno(A.34)  \]
\[\hspace{-2.1cm}\underline{5 + 3} \quad A = \beta^6 \beta^7\beta^8 =A^{-1} = A^\dagger = A^T = A^\ast \eqno(A.35)  \]
\[\hspace{-1.7cm}\underline{4 + 4} \quad A = \beta^5 \beta^6\beta^7 \beta^8 =A^{-1} = A^\dagger = A^T = A^\ast \eqno(A.36)  \]
\[\hspace{-1.3cm}\underline{3 + 5} \quad A = \beta^4 \beta^5 \beta^6\beta^7 \beta^8 =A^{-1} = A^\dagger = A^T = A^\ast \eqno(A.37)  \]
\[\hspace{-2.9cm} 10d: \qquad
B^{(2,4,5,8,9)} = -B^{(2,4,5,8,9)T}\;,  \quad B^{(1,3,6,7,10)}  = B^{(1,3,6,7,10)T} \eqno(A.38) \]
\[\hspace{-2.1cm}C  = B^2B^4B^5B^8B^9 =  C^{-1} =  C^\dagger = -C^T = -C^\ast \eqno(A.39) \]
\[\hspace{-2.4cm}D = B^1B^3B^6B^7B^{10}  = D^{-1} = D^\dagger = D^T = D^\ast \eqno(A40) \]
\[\hspace{-3.2cm}\underline{9 + 1} \quad A = B^{10} = A^{-1} = A^\dagger = A^T = A^\ast \eqno(A.41)  \]
\[\hspace{-1.8cm}\underline{8 + 2} \quad A = B^9B^{10} = -A^{-1} = -A^\dagger = A^T = -A^\ast \eqno(A.42)  \]
\[\hspace{-1.4cm}\underline{7 + 3} \quad A = B^8 B^9B^{10} = -A^{-1} = -A^\dagger = -A^T = A^\ast \eqno(A.43)  \]
\[\hspace{-1.9cm}\underline{6 + 4} \quad A = B^7 B^8B^9 B^{10} =A^{-1} = A^\dagger = A^T = A^\ast \eqno(A.44)  \]
\[\hspace{-1.5cm}\underline{5 + 5} \quad A = B^6 B^7 B^8B^9 B^{10} =A^{-1} = A^\dagger = A^T = A^\ast \eqno(A.45)  \]
\[\hspace{.1cm}\underline{4 + 5} \quad A = B^5 B^6 B^7 B^8B^9 B^{10} = -A^{-1} = -A^\dagger = A^T = -A^\ast \eqno(A.46)  \]
\[\hspace{.8cm}\underline{3 + 7} \quad A = B^4B^5 B^6 B^7 B^8B^9 B^{10} = -A^{-1} = -A^\dagger =- A^T = A^\ast\;. \eqno(A.47)  \]

We have restricted our attention to spaces with at least one time and three spatial dimensions.

We encounter expressions of the form
\[ M^{(d)}_{i\ell} = \left( \tilde{\rho}^\mu \lambda \right)_i \left( \overline{\chi} \tilde{\rho}_\mu \right)_\ell - 
\left( \tilde{\rho}^\mu \chi \right)_i \left( \overline{\lambda} \tilde{\rho}_\mu \right)_\ell \eqno(A.48) \]
in $2 + 1$ and $3 + 1$ and $9 + 1$ dimensions with $\lambda$ and $\chi$ being Majorana spinors in $2 + 1$ and $3 + 1$ dimensions and Majorana-Weyl in $9 + 1$ dimensions.  By using a Fierz transformation $M_{ij}^{(d)}$ can be put into a more suitable form.

In general, in $d$ dimensions, a complete set of $2^{[\frac{d}{2}]} \times 
2^{[\frac{d}{2}]}$ matrices is given by [3]
\[ 1, \;\; \tilde{\rho}^\mu, \;\; \tilde{\rho}^{[\mu_{1}}\tilde{\rho}^{\mu_{2}]} = 
\frac{1}{2!} \left( \tilde{\rho}^{\mu_{1}}\tilde{\rho}^{\mu_{2}} - \tilde{\rho}^{\mu_{2}}\tilde{\rho}^{\mu_{1}} \right),\;\;\tilde{\rho}^{[\mu_{1}}\tilde{\rho}^{\mu_{2}}\tilde{\rho}^{\mu_{3}]} ,\;\; \ldots \tilde{\rho}^{[\mu_{1}} \ldots \tilde{\rho}^{\mu_{d}]} \eqno(A.49) \]
and so in $2 + 1$ dimensions we have $\delta_{ij}, \tilde{\tau}_{ij}^\mu$.  Consequently
\[ \tilde{\tau}_{ij}^\mu \tilde{\tau}_{\mu k\ell} = C_{i\ell} \delta_{kj} + C_{i\ell}^\mu \tilde{\tau}_{\mu kj} \; .\eqno(A.50) \]
Multiplying eq. (A.50) by $\delta_{jk}$ gives
\[ \left( \tilde{\tau}^\mu \tilde{\tau}_\mu\right)_{i\ell} = 2C_{i\ell} \]
or
\[ C_{i\ell} = \frac{3}{2} \delta_{i\ell} \; . \eqno(A.51) \]
Similarly, upon multiplying eq. (A.50) by $\tilde{\tau}_{jk}^\nu$ we obtain 
\[ \left( \tilde{\tau}^\mu\tilde{\tau}^\nu\tilde{\tau}_\mu \right)_{i\ell} = C_{i\ell}^\mu Tr\left( \tilde{\tau}^\mu \tilde{\tau}_\mu \right) \]
or 
\[ C_{i\ell}^\mu = - \frac{1}{2} \tilde{\tau}_{i\ell}^\mu\; . \eqno(A.52) \]
Together, eqs. (A.50-52) show that $M_{ij}^{(3)}$ in eq. (A.48) becomes 
\[\hspace{-2cm} M_{i\ell}^{(3)} = \left( \lambda_j \overline{\chi}_k - \chi_j \overline{\lambda}_k \right) \left( \frac{3}{2} \delta_{i\ell} \delta_{kj} - \frac{1}{2} \tilde{\tau}_{i\ell}^\mu \tilde{\tau}_{ikj} \right) \]
\[ = \frac{3}{2} \delta_{i\ell} \left( - \overline{\chi}\lambda + \overline{\lambda}\chi\right) - \frac{1}{2}\tilde{\tau}_{i\ell}^\mu \left( - \overline{\chi}\tilde{\tau}_\mu \lambda + \overline{\lambda} \tilde{\tau}_\mu \chi \right)\; . \eqno(A.53) \]
In $2 + 1$ dimensions, $\overline{\lambda} = \lambda^\dagger \tau^3$ and $\lambda = C \overline{\lambda}^T$ where  $C = \tau^1 = C^{-1} = C^\dagger = -C^T = -C^\ast$; consequently
\[ \overline{\chi}\lambda = (C^{-1}\chi)^T (C\overline{\lambda}^T) = - \chi^T \overline{\lambda}^T = \overline{\lambda}\chi \eqno(A.54) \]
and
\[ \overline{\chi} \tilde{\tau}^\mu \lambda = -\chi^T C^{-1}
\tilde{\tau}^\mu C\overline{\lambda}^T = - \overline{\lambda} \tilde{\tau}^\mu \chi \; .\eqno(A.55) \]
As a result of eqs. (A. 53-55), we see that
\[ M_{ij}^{(3)} = \tilde{\tau}_{ij}^\mu\; \overline{\chi} \tilde{\tau}_\mu \lambda \;.\eqno(A.56) \]
In $3 + 1$ dimensions, the Fierz expansion takes the form 
\[ \tilde{\gamma}_{ij}^\mu \tilde{\gamma}_{\mu k\ell} = C_{i\ell} \delta_{kj} + C_{i\ell}^\mu \tilde{\gamma}_{\mu kj} + \ldots + C_{i\ell}^{\mu_{1}\ldots \mu_{4}} \tilde{\gamma}_{[\mu_{1}\ldots \mu_{4}]}
\eqno(A.57) \]
Using the properties of $C$ in eq. (A.21), we find that 
\[ \overline{\lambda} \chi = \overline{\chi} \lambda \eqno(A.58) \]
\[ \overline{\lambda}\tilde{\gamma}^\mu   \chi = - \overline{\chi}\tilde{\gamma}^\mu \lambda \eqno(A.59) \]
\[ \overline{\lambda}\tilde{\gamma}^{[\mu_{1}\mu_{2}]} \chi = - \overline{\chi} \tilde{\gamma}^{[\mu_{1}\mu_{2}]} \lambda \eqno(A.60) \]
\[ \overline{\lambda}\tilde{\gamma}^{[\mu_{1}\mu_{2}\mu_{3}]} \chi =  \overline{\chi} \tilde{\gamma}^{[\mu_{1}\mu_{2}\mu_{3}]} \lambda \eqno(A.61) \]
\[ \overline{\lambda}\tilde{\gamma}^{[\mu_{1}\mu_{2}\mu_{3}\mu_{4}]} \chi =  \overline{\chi} \tilde{\gamma}^{[\mu_{1}\mu_{2}\mu_{3}\mu_{4}]} \lambda \eqno(A.62) \]
and thus only $C_{i\ell}^\mu$ and $C_{i\ell}^{[\mu_{1}\mu_{2}]}$ in eq. (A.57) contribute to $M_{i\ell}^{(4)}$ in eq. (A.48).  By multiplying eq. (A.57) with $\tilde{\gamma}_{jk}^\nu$ and $\tilde{\gamma}_{jk}^{[\nu_{1}\nu_{2}]}$, it can be shown that 
\[ C_{i\ell}^\mu = - \frac{1}{2} \tilde{\gamma}_{i\ell}^\mu \eqno(A.63) \]
\[ C_{i\ell}^{\mu\nu} =0 \eqno(A.64)\]
and hence
\[ M_{ij}^{(4)} = (\tilde{\gamma}^\mu)_{ij} \overline{\chi} \tilde{\gamma}_\mu \lambda \eqno(A.65) \]
much like $M_{ij}^{(3)}$ in eq. (A.56).

In $9 + 1$ dimensions, one can use arguments similar to those in $2 + 1$ and $3 + 1$ dimensions. In addition, we have the relation
\[ B^{[A_{1} \ldots A_{9}]} \sim B^{A_{10}} B^{11} \eqno(A.66) \]
and the Weyl property of the spinors so that
\[ B^{11} \lambda = \lambda, \;\; B^{11} \chi = \chi \; .\eqno(A.67) \]
This all results in having
\[ M_{ij}^{(10)} = \left[ \tilde{B}^A \left( \frac{1 + B^{11}}{2}\right)\right]_{ij} \overline{\chi} \tilde{B}_A \lambda \; . \eqno(A.68) \]

In $3 + 3$ dimensions, we need to consider both expressions of the form of eq. (A.48) and of the form
\[ \left(\tilde{\Gamma}^M \lambda\right)_i \left( \overline{\lambda} \tilde{\Gamma}_M\right)_\ell \eqno(A.69) \]
with $\lambda$ and $\chi$ satisfying $\lambda = \lambda_D$, $\chi = \chi_D$ (Majorana condition) and $\lambda = \Gamma^7 \lambda$, $\chi = \Gamma^7 \chi$ (Weyl condition).  It follows that 
\[\overline{\lambda}\chi = - \overline{\lambda}  \chi \;, \quad 
\overline{\lambda} \Gamma^7 \chi = - \overline{\lambda} \Gamma^7  \chi\eqno(A.70a,b) \]
\[ \overline{\lambda} \tilde{\Gamma}^{[M} \tilde{\Gamma}^{N]}\chi = 
- \overline{\lambda} \tilde{\Gamma}^{[M} \tilde{\Gamma}^{N]}\chi\;, \quad 
\overline{\lambda} \tilde{\Gamma}^{[M} \tilde{\Gamma}^{N]} \Gamma^7 \chi = 
- \overline{\lambda} \tilde{\Gamma}^{[M} \tilde{\Gamma}^{N]}\Gamma^7 \chi
\eqno(A.71a,b)\]
\[ \overline{\lambda} \tilde{\Gamma}^{M}\chi = 
 \overline{\chi} \tilde{\Gamma}^{M}\lambda\;, \quad
\overline{\lambda} \tilde{\Gamma}^{M} \Gamma^7\chi = 
 \overline{\chi} \tilde{\Gamma}^{M} \Gamma^7 \lambda \eqno(A.72a,b)\]
 \[\overline{\lambda} \tilde{\Gamma}^{[M} \tilde{\Gamma}^{N} \tilde{\Gamma}^{L]}\chi = - \overline{\chi} \tilde{\Gamma}^{[M} \tilde{\Gamma}^{N} \tilde{\Gamma}^{L]}\lambda\;, \quad
 \overline{\lambda} \tilde{\Gamma}^{[M} \tilde{\Gamma}^{N} \tilde{\Gamma}^{L]} \Gamma^7\chi = - \overline{\chi} \tilde{\Gamma}^{[M} \tilde{\Gamma}^{N} \tilde{\Gamma}^{L]}\Gamma^7\lambda
\eqno(A.73a,b)\]
and so the only relevant terms in the expansion
\[ \tilde{\Gamma}^M_{\;\;\,ij} \;\tilde{\Gamma}_{M\;k\ell} = 
C_{i\ell} \delta_{kj} + C^M_{i\ell} \tilde{\Gamma}_{M\;kj} + 
C_{i\ell}^{M7} \left(\tilde{\Gamma}_M \Gamma^7 \right)_{kj}\]
\[\hspace{-2cm}+ C_{i\ell}^{MN} \left(\tilde{\Gamma}_{[M}\tilde{\Gamma}_{N]}\right)_{kj} \eqno(A.74)\]
\[ +C_{i\ell}^{MN7} \left(\tilde{\Gamma}_{[M}\tilde{\Gamma}_{N]}\Gamma^7\right)_{kj} \]
\[ \hspace{2cm}+C_{i\ell}^{MNL} \left(\tilde{\Gamma}_{[M}\tilde{\Gamma}_N\tilde{\Gamma}_{L]}\right)_{kj} \]
are found to be
\[ C_{i\ell}^M = -2 \tilde{\Gamma}_{i\ell}^M \eqno(A.75) \]
\[ C_{i\ell}^{M7} = 2 \tilde{\Gamma}_{i\ell}^M \Gamma^7 \eqno(A.76) \]
\[\hspace{-1cm} C_{i\ell}^{MNL} = 0\; . \eqno(A.77) \]

\section*{Appendix B}

Some of the conventions we use in the canonical formalism are presented in this appendix.

If $\phi^\mu$ and $\theta_i$ are Bosonic and Fermionic dynamical variables respectively, then when we have a Lagrangian $\mathcal{L}\left( \phi^\mu, \phi^\mu_{,a}; \; \theta_{i} ,\; \theta_{i,a} \right)$ their conjugate momenta are [22] 
\[ p_\mu = \frac{\partial \mathcal{L}}{\partial \phi_{,0}^\mu} \qquad
\pi^i = \frac{\partial \mathcal{L}}{\partial\theta_{i,0}} \; .\eqno(B.1)\]
We use the left derivative for Grassmann variables so that 
\[ \frac{\partial}{\partial\theta^A} (\theta_B \theta_C) = \delta_{AB}\theta_C - \delta_{AC} \theta_B \eqno(B.2) \]
and
\[ \frac{d}{dt} f(\theta(t)) = \dot{\theta}(t) F^\prime (\theta(t)) \; .\eqno(B.3) \]
The Hamiltonian is given by 
\[ \mathcal{H}\left( \phi^\mu, p_\mu ; \theta_i , \pi^i \right) = \phi_{,0}^\mu p_\mu + \theta_{i,0} \pi^i - \mathcal{L}\eqno(B.4) \]
and Poisson Brackets are taken to be
\[ \left\lbrace B_1, B_2 \right\rbrace = \left( B_{1,q} B_{2,p} - B_{2,q} B_{1,p} \right) + \left( B_{1,\psi} B_{2,\pi} - B_{2,\psi} B_{1,\pi} \right) \eqno(B.5) \]
\[ \left\lbrace B, F \right\rbrace = - \left\lbrace F, B \right\rbrace = \left( B_{,q} F_{,p} - F_{,q} B_{,p}\right) + \left( B_{,\psi} F_{,\pi} + F_{,\psi} B_{,\pi} \right) \eqno(B.6) \]
\[ \left\lbrace F_1, F_2 \right\rbrace = \left( F_{1,q} F_{2,p} + F_{2,q} F_{1,p} \right) - \left( F_{1,\psi} F_{2,\pi} - F_{2,\psi} F_{1,\pi} \right) \eqno(B.7) \]
for Bosonic functions $B_i$ and Fermionic functions $F_i$.

\end{document}